# Dual-energy extraction for proton therapy and imaging: validation on a clinical synchrotron-based facility

*Alexander A. Pryanichnikov[a,b,1], Dmitry Shwartz[c], Jon Feldman[d], Joao Seco[b,e,f], Maria Francesca Spadea[a]*

[a] *Institute of Biomedical Engineering, Karlsruhe Institute of Technology (KIT), Fritz-Haber-Weg 1, Karlsruhe, 76131, Germany*

[b] *Division of Biomedical Physics in Radiation Oncology, German Cancer Research Center (DKFZ), Im Neuenheimer Feld 280, Heidelberg, 69120, Germany*

[c] *P-Cure Ltd./Inc, Carmel St. 14, Shilat, 7318800, Israel*

[d] *Sharett Institute of Oncology, Hadassah Medical Center, Hebrew University of Jerusalem, Kalman Ya'Akov Man St., Jerusalem, 91120, Israel*

[e] *Department of Physics and Astronomy, Heidelberg University, Heidelberg, 69120, Germany*

[f] *Heidelberg Institute for Radiation Oncology (HIRO), Heidelberg, 69120, Germany*

[1] Corresponding author at: Institute of Biomedical Engineering, Karlsruhe Institute of Technology (KIT), Fritz-Haber-Weg 1, Karlsruhe, 76131, Germany

Email: alexander.pryanichnikov@kit.edu

## Abstract

**Purpose:** to investigate the feasibility of extracting proton beams with two distinct energies within a single synchrotron cycle. The energy difference between the beams should be sufficient to use one beam as a range probe to guide the second therapeutic beam in a synchrotron-based proton therapy facility.

**Materials and methods:** a clinical synchrotron-based proton therapy facility with a maximum clinical energy of 250 MeV and an experimental energy capability of 330 MeV was utilized. The synchrotron's internal diagnostic equipment, including beam orbit monitors, dipole magnetic field sensors, acceleration frequency monitors, and energy measurement devices, was used to characterize the beams' properties within the synchrotron and during extraction. External dosimetric equipment included a PTW water tank fitted with PTW 34070 and 34080 ionization chambers, as well as an IBA Giraffe ionization chamber array, which were used to independently validate dual-energy extraction at the irradiation isocenter. This equipment was precisely positioned at the isocenter, 70 cm from the accelerator nozzle, using a six-degree-of-freedom robotic positioner.

**Results:** dual-energy extraction within a single synchrotron cycle across three defined energy ranges: low (75–110 MeV), intermediate (120–155 MeV), and high (195–230 MeV) was demonstrated. Both re-acceleration (low to high energy) and de-acceleration (high to low energy) modes were tested and validated.

**Discussion and conclusion:** This preliminary study achieved an energy difference of 35 MeV between extracted beams across three evaluated energy ranges. This corresponds to clinical targets used under standard operating conditions and does not require hardware modifications. Future studies should explore adjusting the control software to enable target switching during a single cycle or the use of suboptimal targets for higher-energy beams. This would allow for active, real-time range probing. Combining modified dual-energy extraction parameters with low-intensity extraction for high-energy beams shows great promise for enabling simultaneous proton imaging and therapy within a single synchrotron cycle.

# 1 Introduction

Image-guided adaptive proton therapy (IGAPT) is an advanced form of proton therapy (PT) used in radiation oncology to treat cancer. IGAPT combines the precision of conventional PT with adaptive treatment planning that accounts for changes in tumor size, shape, and position, as well as variations in surrounding organs at risk (OARs) [1]. Integrating image guidance with appropriate patient immobilization techniques [2] significantly reduces setup uncertainties [3, 4]. This is important for accurately delivering proton therapy, as it is sensitive to deviations due to the sharp dose gradients associated with the Bragg Peak [5].

Currently, image guidance in proton therapy (PT) is predominantly based on X-ray imaging [1,6]. This may include 2D snapshots, cone-beam computed tomography (CBCT), or diagnostic computed tomography (CT) systems located in the treatment room, such as CT-on-rails [7] or upright CTs used in fixed-beam proton therapy (PT) centers [8, 9]. While these X-ray CT systems can effectively track anatomical changes, they expose patients to an additional radiation dose, which can lead to an increased risk of secondary cancers, especially among young patients [10]. Furthermore, X-ray-based treatment planning introduces uncertainties in range when used for proton beam delivery [5]. These uncertainties arise due to the differing physical interactions of X-rays and protons with tissue, with a commonly cited range uncertainty of approximately 3-3.5% [10].

An alternative to X-ray-based imaging is proton imaging [12], which includes proton radiographs (pRad) as a counterpart to planar X-ray images [13, 14] and proton CT (pCT) as a volumetric analogue to CBCT or diagnostic CT [15]. Proton-based imaging eliminates range uncertainties in treatment planning and delivery, as the imaging and treatment modalities share the same particle type and physics behind it [16]. However, implementation during treatment is challenging [17]: therapy requires relatively low-energy protons to stop within the tumor, while imaging demands high-energy protons to traverse the patient's body.

An elegant solution has been demonstrated in ion therapy using two ion species with similar mass-to-charge ratios, where the lighter ion serves as a range probe for the heavier one [18,19,20,21]. However, this solution is not applicable to standard PT because only one particle species, protons, is available. A solution for proton therapy can be rapidly switching between two proton energies within the accelerator or beam delivery system. This is particularly challenging in cyclotron and synchrocyclotron systems, which are currently the most common types of accelerators for protons, due to the mechanical components involved in adjusting beam energy (e.g., energy degraders placed in the beamline). Nevertheless, recent studies have shown promising progress in reducing energy-switching times, even in these systems [22,23,24].

In contrast, synchrotrons and linear accelerators can adjust beam energy through electromagnetic means. Among these, synchrotrons are already used clinically in several PT centers and are gaining broader adoption. Synchrotrons are cyclic accelerators that typically produce a single proton beam energy per cycle, with each cycle lasting several seconds. A promising approach is to extract multiple energies within a single cycle—first for imaging and then for treatment or vice versa. Multi-energy extraction is not a new concept [25]; it has been explored previously and even implemented clinically [26], though primarily to reduce treatment time by limiting the number of cycles required [27]. In such applications, only small energy changes are typically needed.

In this paper, we investigate a novel dual-energy extraction mode within a single synchrotron cycle for the clinical facility. The objectives of this study were threefold: (1) to demonstrate the feasibility of dual-energy extraction with a significant energy difference suitable for range probing, using a commercially available synchrotron-based proton therapy system; (2) to explore the limitations of this dual-energy mode; and (3) to identify the system modifications needed to make this mode viable for real-world clinical applications.

## 2 Materials and Methods

*Proton synchrotron and internal equipment*

The synchrotron used in this study is part of the P-Cure Proton Therapy Solution (Shilat, Israel) [28]. The system includes a compact, weak-focusing synchrotron for proton acceleration, a beam delivery system, and a patient positioning and X-ray imaging system (upright CT and planar X-ray radiographs)—all located within a single vault to enable gantry-less operation and upright patient treatment.

In clinical mode, the synchrotron delivers protons across a water-equivalent range of 2 to 38 cm, corresponding to 2100 discrete energy layers between 50 and 250 MeV. For research applications, the energy can be increased up to 330 MeV, enabling a proton range in water of up to 60 cm—suitable for full-body proton imaging. The system features a fixed-beam pencil beam scanning (PBS) delivery with a maximum field size of 30 cm × 28 cm at an isocenter located 70 cm from the nozzle. The transverse spot size at isocenter is energy-dependent, reaching a minimum $\sigma_x = \sigma_y = 1.9$ mm at 330 MeV.

The accelerator operates via an on-demand sequence: an injection system generates 1.1 MeV protons, which are injected in a single turn into the synchrotron. The beam is captured within ~2 ms and accelerated to the desired energy with a precision of ±0.1 MeV and a ramp rate of approximately 275 MeV/s. After acceleration, particles can be extracted from the synchrotron for 5 ms to several seconds, depending on the number of spots and beam intensity required by the irradiation or treatment plan. Extraction occurs via slow resonant radio frequency knock out (RF-KO) scheme with the use of thin beryllium extraction target. Prior to extraction one of four targets is moved from parking to work position and in addition the beam closed orbit is distorted by steering coils in order to shift the beam closer to electrostatic septum. Weak transverse RF excitation forces small fraction of the beam to abruptly lose the energy while hitting the target and to escape the ring via electrostatic and magnetic septa.

Multi-energy (or currently dual-energy) extraction is available under an R&D-specific operational mode, which enables partial beam extraction, re-acceleration (or deceleration), and subsequent second extraction within the same synchrotron cycle. This is accomplished via dedicated waveforms uploaded to the magnet power supplies, RF system, extraction elements, and beamline correctors. This mode bypasses some clinical interlocks and allows for manual control over key parameters. These beam parameters were verified using internal instrumentation: (1) Hall sensors in each dipole for real-time (every 0.2 ms) magnetic field monitoring, (2) RF system readouts for energy validation, (3) position sensors for extraction targets; as well as the dose monitoring system (DMS) instrumentation such as (6) an optical scintillator-based extracted beam current monitor and (5) an ionization chamber array.

The equipment in the treatment room is represented by the Patient Robotic Positioning and Imaging System (P-ARTIS), which consists of: (1) a patient positioning system (PPS) with a patient chair or deck; (2) an upright 4-dimensional computed tomography (4DCT) system, which is used for treatment planning, simulation, and verification of patient positioning prior to treatment, as well as for treatment adaptation if necessary; and (3) an orthogonal 2D kilovoltage (kV) imaging system, which is used solely for verification of patient positioning in the treatment isocenter. For this study, the PPS was used to place external dosimetric detectors (Figure 1 A and B) that have been calibrated for loads up to 180 kg with ±0.5 mm accuracy (95% confidence).

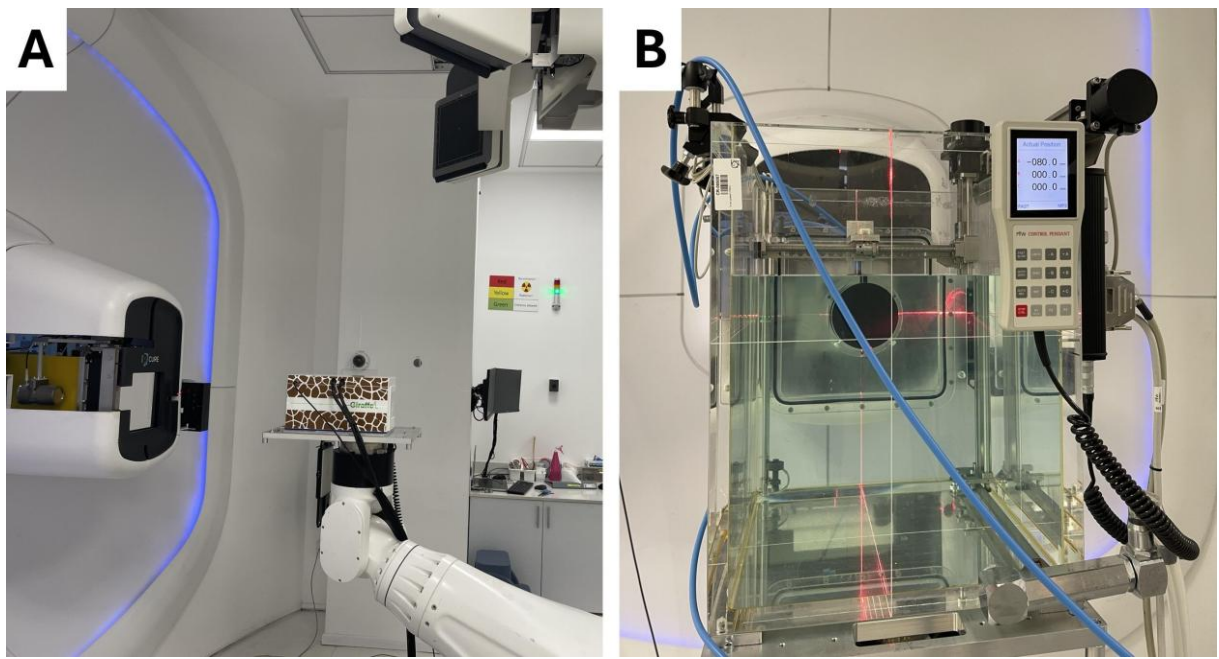

Figure 1: External dosimetry setup: (A) A multi-layer ionization chamber and (B) A water tank containing two ionization chambers were placed in the irradiation isocenter for dual-energy extraction verification.

*External dosimetry*

To independently validate the dual-energy extraction capability of the synchrotron, integrated depth-dose curves (IDDs) were acquired at the irradiation isocenter using two clinically certified dosimetry systems suitable for proton beam measurements.

The first system employed was the IBA Giraffe (IBA Dosimetry, Schwarzenbruck, Germany), a two-dimensional array of parallel-plate ionization chambers designed for high-resolution proton dosimetry. The Giraffe consists of 180 chambers arranged in a linear geometry, allowing for simultaneous acquisition of dose profiles at multiple depths. The data were analyzed using IBA myQA software, which provides depth-dose profiles and range parameters with submillimeter resolution.

The second system used for IDD acquisition consisted of two parallel-plate ionization chambers, the PTW Bragg Peak Chamber 34070 and 34080 (PTW, Freiburg, Germany), combined with a PTW Tandem XDR electrometer and a PTW MP3 water phantom scanning system. This setup enables high-precision measurements of pencil beam IDDs across a broad range of proton energies. The chambers were calibrated in terms of absorbed dose to water and scanned through the proton beam to measure dose deposition as a function of depth. The chamber's movement was synchronized with the collected charge.

*Study design*

To demonstrate the feasibility of dual-energy extraction for applications such as range probing or proton imaging, the following experimental protocol was implemented:

1. Energy range selection: Three representative energy ranges were selected—low, intermediate, and high.

2. Energy pairing: Within each range, two discrete energies with the maximum feasible energy separation were identified for dual-energy extraction experiments.

3. Experimental scenarios: Measurements were performed under six distinct scenarios: (1) Single-energy extraction: (a) first cycle of high energy only; (b) followed by the second cycle of low energy only; (2) Dual-energy extraction (low → high): both energies extracted sequentially within a single accelerator cycle; (3) Dual-energy extraction (high → low): reverse sequence within a single cycle. Each of these three scenarios was tested under two accelerator operation modes: (I) *Fixed cycle:* the duration of the cycle is predefined, independent of beam extraction; (II) *Free cycle:* the cycle automatically terminates after the requested proton dose is fully extracted

4. Spot Delivery Parameters: For each scenario, two spills were delivered per energy, targeting a single central spot at coordinates (0 mm, 0 mm). The planned beam intensities were set to $10^7$ protons for high-energy beams and $10^8$ protons for low-energy beams.

All measurements were conducted using an anti-hysteresis cycle, in which the synchrotron's magnetic field was ramped up to the level corresponding to 250 MeV at the end of each cycle. This approach was necessary to ensure stable and reproducible accelerator performance throughout the experiments.

## 3 Results

*Internal measurements*

Before conducting the main experiments, limitations of the dual-energy extraction process were evaluated with respect to the current control software and accelerator hardware.

Two primary constraints were identified: (1) Single extraction target per cycle: due to control sequence limitations, it is only possible to use a single extraction target within one accelerator cycle. This restriction splits the energy range into non-interleaved regions: 70-115 MeV, 115-185 MeV and 185-250 MeV. (2) Limited number of points in the waveform table sent to the main power supply digital-to-analog converter (DAC). This restricts the number of energy transitions and the smoothness of acceleration/deceleration curves. Nevertheless, it was determined that an energy switch of up to $\Delta E = 35$ MeV is achievable across all three energy ranges tested as shown in Table 1.

Table 1. Selected Energy Pairs for Validation Measurements

| Energy range | Bottom energy, MeV | Top energy, MeV |
|---|---|---|
| Low | 75 | 110 |
| Intermediate | 120 | 155 |
| High | 195 | 230 |

Following this determination, a series of measurements (for energies in Table 1) was performed according to the experimental protocol described in the Study design section. Internal parameters—including magnetic field and accelerating RF frequency—were recorded for selected energy pairs under different cycle conditions: *fixed* 3-second cycles for low → high switching, and *fixed 3-second* cycles for high → low switching as presented in Figure 2 and Figure 3, respectively.

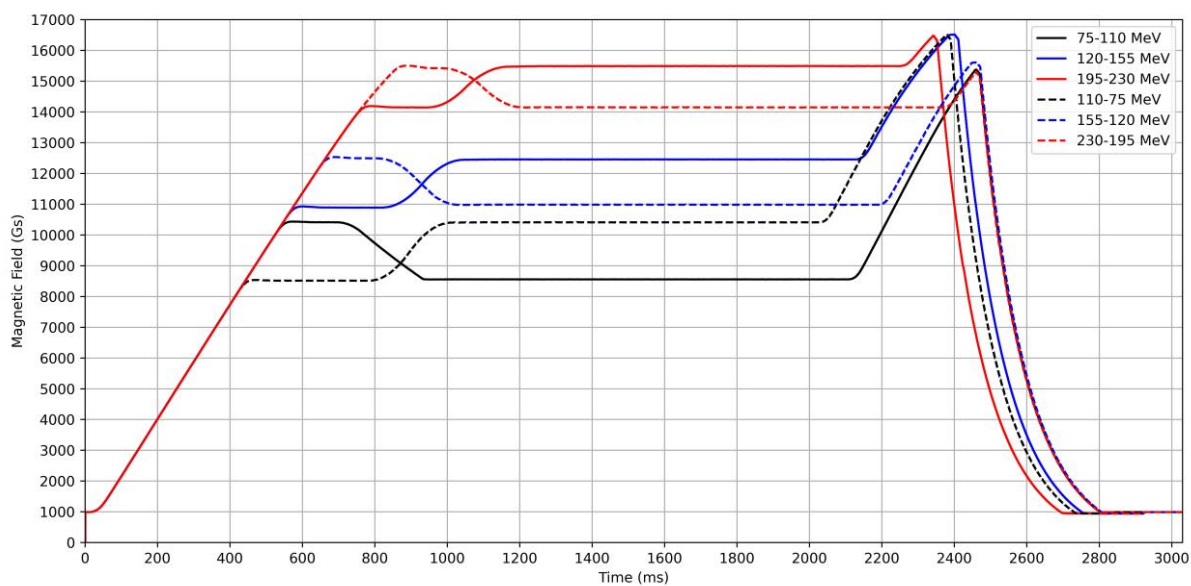

Figure 2: An example of magnetic field measurements taken from single Hall sensors installed in synchrotron dipole magnet #11 (the temporal resolution is 0.2 ms).

As shown in Figure 2, the magnetic field profile comprises 12 distinct segments: (1) Field stabilization at injection; (2) Smooth ramp to the acceleration phase; (3) Beam acceleration; (4) Smooth transition to the extraction plateau (first energy); (5) Extraction plateau for the first energy; (6) Smooth ramp to the second acceleration phase; (7) Beam re-acceleration; (8) Smooth ramp to the second plateau; (9) Extraction plateau for the second energy; (10) Magnetic field ramp-up to 250 MeV equivalent (anti-hysteresis cycle); (11) Ramp-down to a field below injection level (12) Final field stabilization at injection level. The main limitation for switching energies currently lies in the number of interpolation points available for the smooth transitions required between plateaus. These transitions must be finely controlled to minimize beam losses.

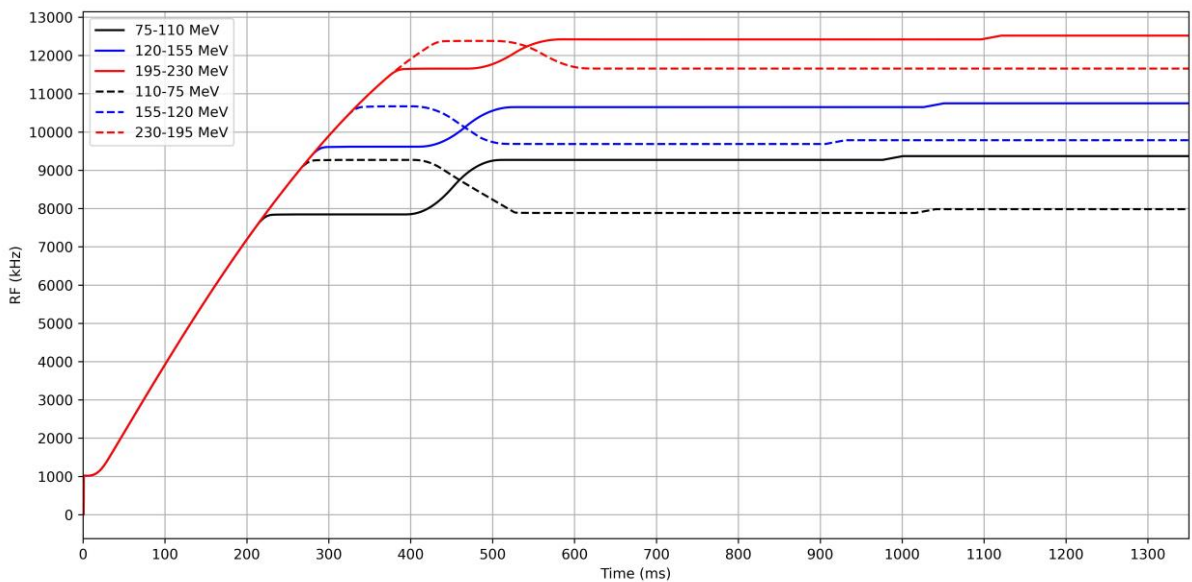

Figure 3: Accelerating RF frequency measurements during energy transitions (the temporal resolution is 0.2 ms).

All "low → high" transitions were performed using *fixed* 3-second cycles; all "high → low" transitions used *free* cycles. Each cycle included an anti-hysteresis field ramp.

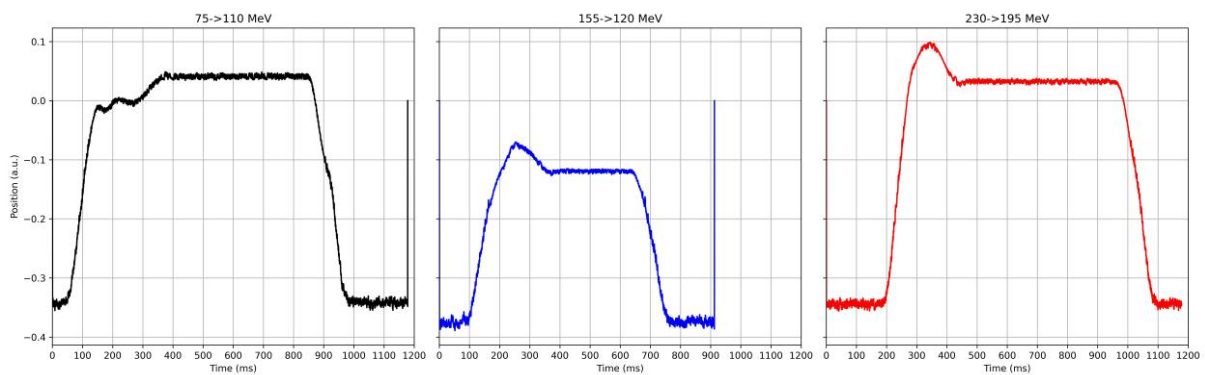

Figure 4: Extraction target positioning within the synchrotron ring (the temporal resolution is 0.2 ms).

As was mentioned above only one extraction target can be set per accelerator cycle. Figure 4 illustrates the target position control. The active target corresponds to the first energy selected in a dual-energy sequence. For example: in the low energy range, a 100 µm thickness target was used when switching from low to high energy (Figure 4 left) while a 200 µm and 400 µm were used in the intermediate and high energy ranges correspondingly (Figure 4 center and right). In the current study we didn't cross target-change energies, nevertheless it is possible even without control system upgrade. This just means that the extraction settings for the second energy will be suboptimal, as each energy requires dedicated target thickness to achieve optimal beam extraction efficiency.

*External measurements*

The primary objective of this study was to demonstrate the feasibility and reproducibility of dual-energy extraction with a significant energy difference. To validate this, external dosimetric measurements were performed using the equipment described in the External dosimetry section.

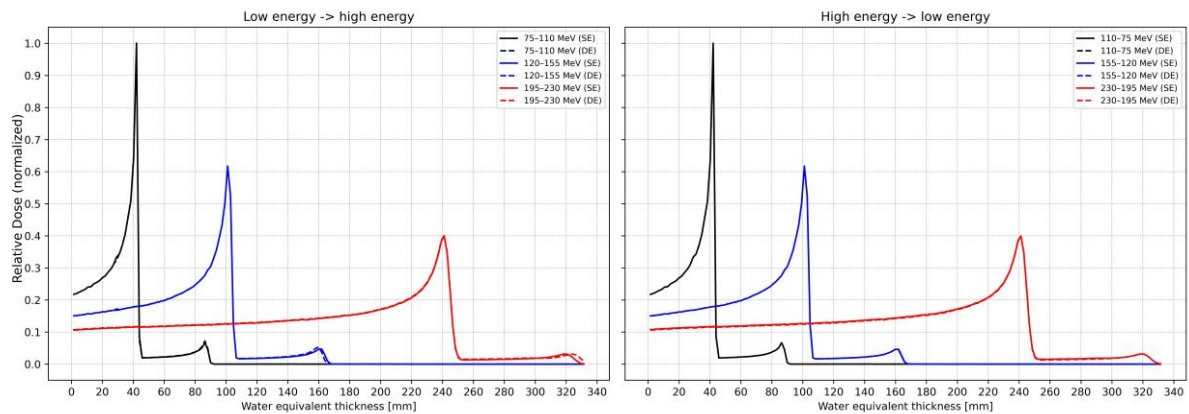

Figure 5: Comparison of single and dual-energy extraction using a two-dimensional array of parallel-plate ionization chambers (IBA Giraffe) for both low-to-high and high-to-low energy extraction modes.

Figure 5 shows the IDD curves normalized to the 75 MeV peak obtained with the IBA Giraffe system for single- and dual-energy extraction modes. In the dual-energy scenario, two distinct energy layers were extracted with a single accelerator cycle, whereas two separate accelerator cycles were used in the single-energy configuration. The results demonstrate good reproducibility and stability in both "high-to-low" and "low-to-high" switching directions, even with a 10:1 intensity ratio between the two energy layers. There was no significant distortion or dose inconsistency, confirming the stability of the dual-energy operation.

The PTW Bragg Peak chamber measurements require individual scans at each depth and were validated only for the low-to-high energy switch configuration. Figure 6 shows the results, demonstrating submillimeter agreement (within 1 mm) with the IBA Giraffe data and planned range values.

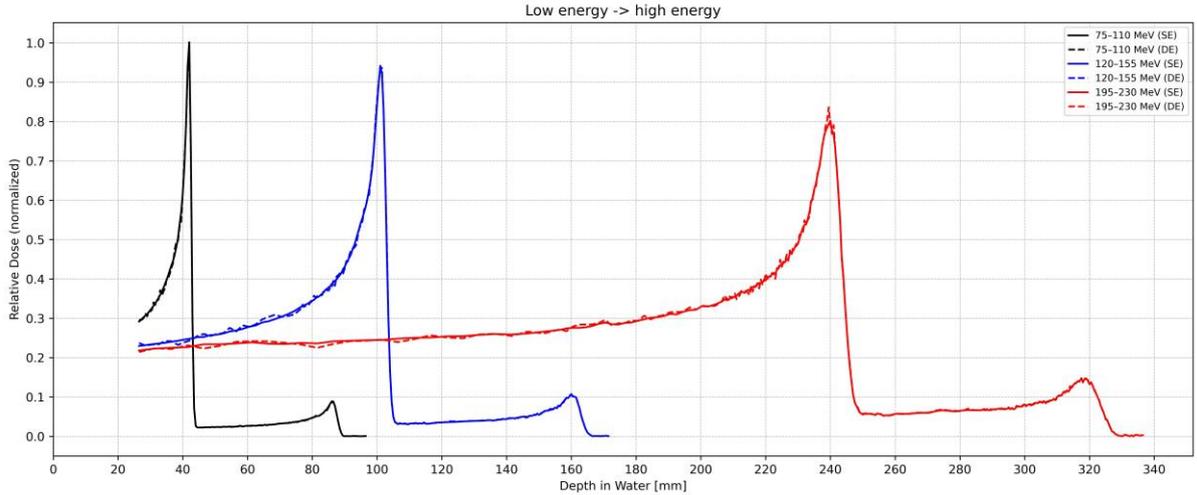

Figure 6: Comparison of single and dual-energy extraction using two parallel-plate ionization chambers (PTW 34070 and 34080) in a water phantom, for the low-to-high energy configuration.

These external dosimetric measurements confirm the technical feasibility and accuracy of dual-energy extraction without significant degradation in dose distribution or extraction consistency.

## 4 Discussion

This study presents the first experimental demonstration of two-energy extraction with a significant energy difference from a medical proton synchrotron within a single accelerator cycle. The ability to deliver two distinct energy layers spanning up to a 35-MeV difference without modifying the existing accelerator hardware or control software represents an important technological advancement. This capability opens new possibilities in proton therapy, particularly for range probing and proton imaging applications, where rapid and significant energy modulation is essential. The consistency observed in IDDs across different switching sequences and operational modes, even with a 10:1 beam intensity ratio, suggests that dual-energy operation can be implemented without compromising dosimetric accuracy or system stability.

As this was a proof-of-concept study, several limitations must be acknowledged before future technology implementation can be considered. One major area requiring improvement is the beam intensity for the imaging or range-probing component. For imaging, especially with proton radiography or CT, the beam intensity should be reduced significantly. Prior work with similar synchrotrons [29] has shown that ultra-low extraction intensities, on the order of $10^4$-$10^5$ protons per second, are achievable through specific extraction tuning [30,31,32]. These settings are not yet implemented in the current configuration but could be introduced in R&D mode, potentially enabling intensity differences exceeding a factor of $10^3$ between treatment and imaging beams. Future research should explore the integration of such low-intensity extraction settings into dual-energy operation.

To enable proton imaging during dual-energy extraction, the system must also support beam scanning. Currently, this is not possible because the control software does not permit updated quadrupole settings within the cycle. Updating the software to allow for multiple settings for quadrupole lenses and beam optics during dual-energy cycles would be a necessary step in expanding imaging capabilities. Additionally, delivering individual spots in less than the current minimum of 5 ms is required. Recent studies have demonstrated the feasibility of high-resolution proton imaging using spot durations as low as 1 ms over large fields (up to 12 cm × 12 cm), which could cover a substantial number of treatment layers [33]. Thus, enabling shorter spot durations in dual-energy mode should be prioritized in future system development.

There are also two technical constraints that were previously discussed and that merit further consideration. First, only one extraction target can be used per cycle. While this does not significantly impact imaging performance, as lower efficiency is acceptable for range probing or imaging, it may affect the quality or intensity of the subsequent therapy beam. This is especially true when switching from high to low energy. To ensure clinical beam integrity, use the target appropriate for the treatment energy. Second, the current main power supply memory capacity for magnetic field control limits the number and granularity of possible energy transitions. Hardware upgrade would allow for larger energy separations and more complex switching profiles, which could be useful for extended imaging studies or future clinical applications.

This study did not focus on quantifying beam losses during energy switching especially in high-to-low direction. The current power supply was not designed to decrease magnetic field in a well-controlled way instead evacuating the stored energy into external shunt. Another technical difficulty come from the non-standard hysteresis loops. However, initial observations suggest that switching from low to high energy could improve treatment outcomes. In this configuration, the high-energy layer used for imaging could verify the dose delivered by the preceding treatment layer or predict the dose to be delivered by subsequent layers.

Despite these limitations, the results demonstrate the feasibility of dual-energy extraction suitable for the range probing or proton imaging using a commercial clinical synchrotron. The system demonstrated stable operation, consistent beam parameters, and precise dose delivery throughout energy transitions. These findings lay the technical groundwork for the further development of proton-based imaging and range verification technologies, as well as the integration of these capabilities into clinical workflows.

## 5   Conclusion

This study provides the first experimental validation of dual-energy proton extraction within a single cycle of a clinical, synchrotron-based therapy system. The study demonstrates the technical feasibility of delivering two significantly different energy layers ($\Delta E = 35$ MeV) without modifying the hardware or software. Consistency in beam quality and dose distribution was confirmed through internal instrumentation and external dosimetry.

These results highlight the potential of this approach for advanced applications, such as proton imaging and in vivo range verification. However, several technical limitations remain, including constraints in the control software, limits in spot duration, the use of a single extraction target, and combination with low-intensity extraction. However, these challenges can be overcome with targeted system upgrades. These results lay the groundwork for future research and development aimed at integrating dual-energy functionality into proton therapy and imaging workflows.

**Funding Statement**

This study has not received any financial support.